\documentclass[runningheads]{llncs}
\usepackage[T1]{fontenc}

\usepackage{mathtools}
\usepackage{graphicx}

\usepackage{amsmath}
\usepackage{bm}

\usepackage[T1]{fontenc}
\usepackage{multirow}
\usepackage{graphicx}
\usepackage{amssymb}
\usepackage{amsfonts}
\usepackage{booktabs}
\usepackage{array}

\newcolumntype{L}[1]{>{\raggedright\let\newline\\\arraybackslash\hspace{0pt}}m{#1}}
\newcolumntype{C}[1]{>{\centering\let\newline\\\arraybackslash\hspace{0pt}}m{#1}}
\newcolumntype{R}[1]{>{\raggedleft\let\newline\\\arraybackslash\hspace{0pt}}m{#1}}

\makeatletter
\newcommand{\thickhline}{%
    \noalign {\ifnum 0=`}\fi \hrule height 1pt
    \futurelet \reserved@a \@xhline
}
\newcolumntype{"}{@{\hskip\tabcolsep\vrule width 1pt\hskip\tabcolsep}}
\makeatother

\begin{document}

\title{TBSS++: A novel computational method for Tract-Based Spatial Statistics}
\titlerunning{Deep-TBSS}

\author{Davood Karimi\inst{1}, Hamza Kebiri\inst{1,2}, and Ali Gholipour\inst{1}}

%
\institute{Computational Radiology Laboratory (CRL), Department of Radiology, Boston Children's Hospital, and Harvard Medical School, USA \and Department of Radiology, Lausanne University Hospital (CHUV) and University of Lausanne (UNIL), Lausanne, Switzerland}%

%
\maketitle              

\begin{abstract}

Diffusion-weighted magnetic resonance imaging (dMRI) is widely used to assess the brain white matter. One of the most common computations in dMRI involves cross-subject tract-specific analysis, whereby dMRI-derived biomarkers are compared between cohorts of subjects. The accuracy and reliability of these studies hinges on the ability to compare precisely the same white matter tracts across subjects. This is an intricate and error-prone computation. Existing computational methods such as Tract-Based Spatial Statistics (TBSS) suffer from a host of shortcomings and limitations that can seriously undermine the validity of the results. We present a new computational framework that overcomes the limitations of existing methods via (i) accurate segmentation of the tracts, and (ii) precise registration of data from different subjects/scans. The registration is based on fiber orientation distributions. To further improve the alignment of cross-subject data, we create detailed atlases of white matter tracts. These atlases serve as an unbiased reference space where the data from all subjects is registered for comparison. Extensive evaluations show that, compared with TBSS, our proposed framework offers significantly higher reproducibility and robustness to data perturbations. Our method promises a drastic improvement in accuracy and reproducibility of cross-subject dMRI studies that are routinely used in neuroscience and medical research.


\end{abstract}

\section{Introduction}

Diffusion of water in the brain tissue is influenced by tissue micro-structure. Diffusion MRI (dMRI) methods can characterize this process quantitatively \cite{le2001diffusion,basser1994mr}. These methods are widely used in medicine and neuroscience. Thousands of studies have used dMRI parameters such as fractional anisotropy (FA) as biomarkers to assess normal brain development and to characterize the impacts of various neurological diseases \cite{assaf2008diffusion,shenton2012review,lebel2019review}.

\emph{Tract-specific studies} analyze dMRI-computed biomarkers on specific white matter tracts \cite{smith2014cross}. They may consider the same subject(s) over time (longitudinal studies) or compare different cohorts of subjects such as healthy and diseased (population studies). They are among the most common and most fruitful classes of dMRI studies because many neurodevelopmental deficits and neurological disorders are linked to damage to the tissue micro-structure on specific tracts \cite{pini2016brain,sexton2011meta,zhuang2010white}. The success of these studies depends on precise alignment of different brain scans to ensure that the same tracts are compared between scans/subjects. This is a complex and highly error-prone computation. Existing computational methods vary greatly in terms of spatial accuracy and the required manual work \cite{smith2014cross}. For example, voxel-based morphometry methods are simple, but they are incapable of accurate tract-specific comparisons \cite{ashburner2000voxel}. There are also semi-automatic and automatic tractography-based methods that require performing tractography on each scan \cite{pagani2005method,suarez2012automated}.

Currently, Tract-Based Spatial Statistics (TBSS) \cite{smith2006tract} is the \emph{de-facto} method for tract-specific dMRI analysis. It can analyze the entire brain and consists of five steps: (1) Compute FA images for each subject; (2) Register the FA images to a common space; (3) Compute the voxel-wise average of the registered FA images and create the skeleton of this mean FA; (4) Project subjects' FA images onto the skeleton; (5) Perform statistical analysis across subjects on the skeleton. Despite its popularity, TBSS suffers from important shortcomings that have been extensively reported. Here we briefly list some of them: (1) FA-based registration is inaccurate, and subsequent steps in TBSS fail to correct the registration errors \cite{zalesky2011moderating}. (2) Presence of lesions reduce the accuracy of TBSS, which is a serious limitation because such pathology may be the main focus of the analysis \cite{jones2010twenty}. (3) Results are sensitive to measurement noise, choice of FA template, and FA threshold \cite{madhyastha2014longitudinal}. (4) TBSS is prone to confusing adjacent tracts even under ideal and simulated settings \cite{bach2014methodological}. (5) Its central assumption is that the locations of interest are the tract centers, defined as the voxels with the highest FA. This assumption may not be valid. (6) Computational steps such as skeletonization and skeleton projection can introduce systematic bias \cite{edden2011spatial}. (7) TBSS is not suited for all tract shapes and may produce erroneous results on small tubular tracts \cite{bach2014methodological}. (8) Although TBSS is an automatic method, it involves several settings that may influence the results. The impacts of these settings are not fully clear, and one should visually verify the results of different processing steps \cite{madhyastha2014longitudinal}.

These limitations significantly compromise the validity of the results of cross-subject dMRI studies. It has been shown that tweaking the settings of the existing methods may lead to completely different conclusions \cite{bach2014methodological,keihaninejad2012importance}. Alternative methods have been proposed to improve certain aspects of TBSS \cite{yushkevich2009structure,yushkevich2007structure}. However, these alternative methods have their own limitations and none of them has reached TBSS's popularity. Despite years of effort, existing methods suffer from serious shortcomings that are data-dependent and difficult to predict.

Given the increasing popularity of dMRI and the growing size of datasets, there is an urgent need for accurate and reliable computational tools. The goal of this paper is to develop and validate a new computational framework, in large part based on deep neural networks, that overcomes some of the limitations of the existing methods. We show that the new methods enable unprecedented accuracy in tract-specific cross-subject analysis of dMRI data. Therefore, they can have a significant impact as they dramatically enhance the reliability and reproducibility of longitudinal and population studies with dMRI.

\section{Materials and methods}

\subsection{Data}

We use two datasets in this work. (1) Human Connectome Project (HCP) dataset. We use 105 adult brains from the HCP dataset \cite{van2013wu} and manual segmentations of 72 tracts on these brains available via \cite{wasserthal2018tractseg}. The dMRI scans in this dataset include high angular resolution multi-shell (b=1000, 2000, and 3000) measurements. In all experiments with this data, we used 70 subjects for model development and the remaining 35 for test. (2) Calgary Preschool dataset \cite{reynolds2020calgary,reynolds2019global}, which we only use for further validation. This dataset includes 396 dMRI scans from 120 children 2-8 years of age. Each scan in this dataset has 35 single-shell (b=750) measurements.

\subsection{Methods}

\begin{figure*}[!htb]
\centering
\includegraphics[width=0.99\textwidth]{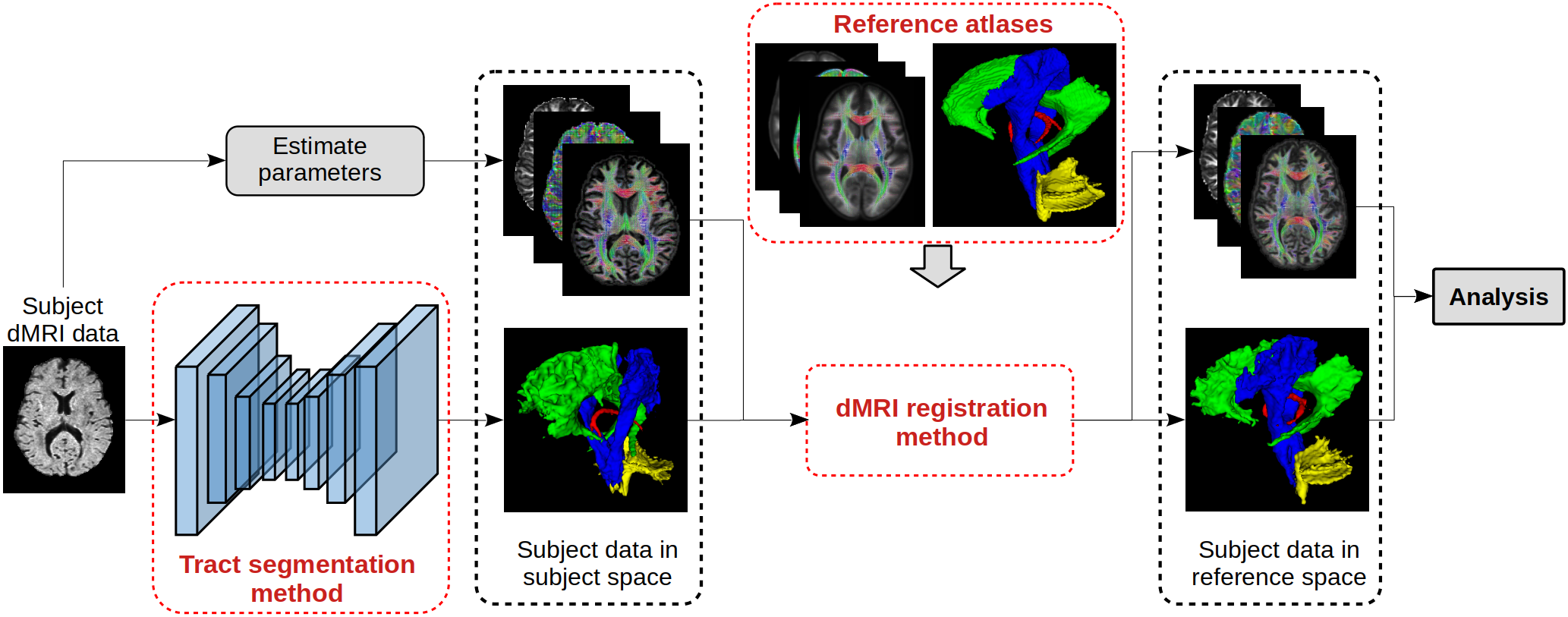}
\caption{\small{A graphical summary of the proposed computational framework. To simplify the illustration, we have shown only one subject and only four tracts.}}
\label{fig:framework_overview}
\end{figure*}

Figure \ref{fig:framework_overview} shows the proposed computational framework. It consists of three main components that we describe in the following three sub-sections.

\subsubsection{Tract segmentation method.} 

Many cross-subject analysis methods, including TBSS, do not segment individual tracts, which causes two problems: (1) Tract-specific computation is not possible. In order to conduct a tract-specific analysis, tracts will have to be segmented either manually or with an auxiliary method. (2) Ambiguity in the location and extent of the tracts can result in significant errors (e.g., mis-assignment of voxels) even for prominent tracts such as corpus callosum and cingulum \cite{bach2014methodological}. Segmentation of individual tracts can overcome this fundamental limitation. To this end, we use a fully-convolutional network similar to nnU-Net \cite{isensee2021nnu} to segment the individual tracts. The segmentation method itself is not a novel aspect of this work. In fact, similar methods have been proposed in recent works \cite{wasserthal2018tractseg}.

\subsubsection{Atlases of white matter tracts.} 

We computed fiber orientation distribution images for the 70 training subjects using the MSMT-CSD method \cite{jeurissen2014multi}. We aligned all 70 brains into a common space using an iterative succession of rigid, affine, and non-linear registrations computed based on fiber orientation distributions using the registration method proposed in \cite{raffelt2011symmetric}. We then used the composite deformation fields for each subject to align individual tract segmentations into the same space and used voxel-wise averaging and thresholding at 0.50 to compute tract atlases. We further applied the computed deformation fields to the diffusion MRI data of each subject, represented in spherical harmonics basis, to compute a corresponding atlas. These atlases would serve as an unbiased reference space where we could register the data from individual subjects for tract-specific analysis.

\subsubsection{dMRI registration method.} 

Accurate registration of brain dMRI data from different subjects is the most central computation in cross-subject studies. Existing methods mostly rely on registration based on fractional anisotropy (FA) images, which can be highly inaccurate. In our proposed framework, we require a flexible registration technique to register the subject brains to the reference atlas. Specifically, similar to our atlas computation described above, we use the method of \cite{raffelt2011symmetric} that is based on computing a nonlinear deformation map based on fiber orientation distributions.

\textbf{To summarize:} the combination of the tract segmentation method, the atlas, and the registration method constitute our proposed framework. The method works as follows (Figure \ref{fig:framework_overview}): (i) Compute the parameter(s) of interest (e.g., FA) and segment tracts of interest for every subject; (ii) Align all subjects into the reference atlas space; (iii) Perform statistical analysis between desired sub-groups of the subjects to determine the differences in the parameters on each tract.

\section{Results and Discussion}

\subsection{Assessment of the three components of the framework}

Our main interest is on the effectiveness of the entire framework for performing tract-specific studies. Nonetheless, we first briefly explain the evaluation of the three components.

\subsubsection{Assessment of the segmentation method.} 

Our deep learning based segmentation method achieved a DSC of $0.826 \pm 0.076$, which is comparable with the state of the art method \cite{wasserthal2018tractseg} on the same dataset. However, unlike \cite{wasserthal2018tractseg} that needs the data in all three shells, our method requires only the measurements in the b=1000 shell.

\subsubsection{Assessment of atlases of white matter tract.} 

\begin{figure*}[!htb]
\centering
\includegraphics[width=1.0\textwidth]{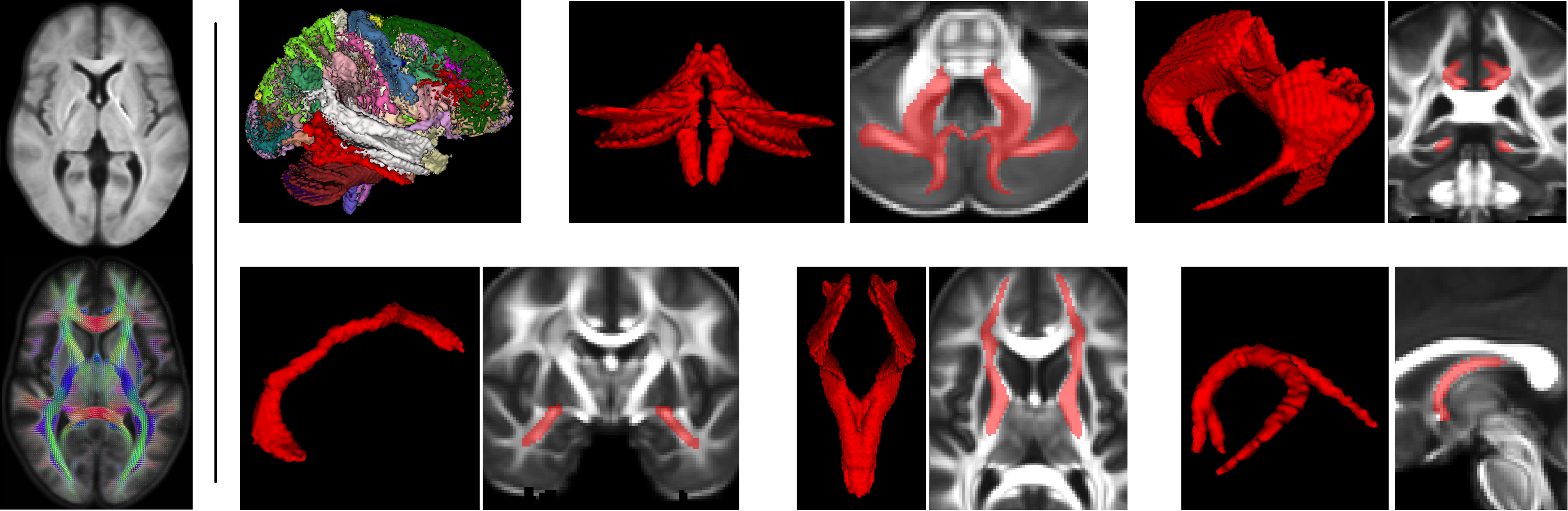}
\caption{\small{LEFT: one slice from the dMRI data and fiber orientation distribution of the generated atlas. RIGHT (from top-left to bottom-right): a volume showing all 72 tracts, inferior cerebellar peduncle, cingulum, anterior commissure, frontopontine tract, and fornix.}}
\label{fig:atlas_examples}
\end{figure*}

Figure \ref{fig:atlas_examples} shows several views from this atlas. All tracts were reconstructed in the atlas with high detail. In order to quantitatively assess the accuracy of the atlas, we divided the 70 training subjects into two groups, each with 35 subjects, generated two separate atlases, and computed the DSC between white matter tracts in the two atlas. The mean DSC across all tracts was 0.962, suggesting that the atlases could be used as an unbiased reference space for registering and comparing data from individual subjects.

\subsubsection{Assessment of the registration method.} The most common approach to quantitative evaluation of registration accuracy is in terms of the overlap of corresponding structures between moving and fixed images \cite{dalca2019unsupervised,zhao2019recursive}. Following this approach, we assessed registration accuracy in terms of DSC and average surface distance (ASD) between the warped subject tracts and atlas tracts. For this evaluation, we considered four of the tracts that spanned most of the brain volume: thalamo-prefrontal, striato-parietal, and anterior midbody and isthmus of corpus calosum. The registration accuracy in terms of DSC was $0.785 \pm 0.074$, while the ASD was $0.775 \pm 0.082$ mm.

\subsection{Assessment of the overall framework for tract-specific cross-subject analysis}

In general, it is difficult to compare different computational frameworks for cross-subject studies. The aim of such studies is to compare two or more groups of subjects and identify the ``true'' differences between the groups. However, discovering such differences, which typically have complex biological underpinnings, is the \emph{goal} of these studies and they are not known in advance. Therefore, for assessing computational methods, it is common to use synthetic data, simulations, and test-retest/reproducibility experiments \cite{bach2014methodological,smith2006tract,schwarz2014improved,arkesteijn2019orientation,madhyastha2014longitudinal}. Following similar evaluation strategies, we applied our new framework in several experiments and compared it with the official implementation of TBSS \cite{TBSS_official}. In these experiments, described below, to enable tract-specific comparison of our method with TBSS, we non-linearly registered the mean FA image from TBSS to the atlas FA image.

\textbf{Reproducibility.} Each dMRI scan from the HCP dataset has 60-90 measurements in the b=1000 shell. For each of the 35 test subjects from this dataset, we selected two independent subsets of 30 measurements from this shell. We applied our method and TBSS on these two measurement subsets and computed the average FA on each tract for each subject. Let us denote the absolute difference between the mean FA for tract $i$ and subject $j$ with $\Delta \text{FA}_{\text{tract\_}i,j}= \big| \text{FA}_{\text{tract\_}i,j}^{\text{subset\_2}} - \text{FA}_{\text{tract\_}i,j}^{\text{subset\_1}} \big|$. Ideally, $\Delta \text{FA}_{\text{tract\_}i,j}$ should be zero. Let us denote the average of $\Delta \text{FA}_{\text{tract\_}i,j}$ across all subjects with $\Delta \text{FA}_{\text{tract\_}i}$. We computed $\Delta \text{FA}_{\text{tract\_}i}$ on all tracts for our method and TBSS. On 35 out of 41 tracts, $\Delta \text{FA}_{\text{tract\_}i}$ was smaller for our method than for TBSS. On 25 of these 35 tracts, the difference was statistically significant at $p=0.01$ computed using paired t-tests. On the remaining 6 tracts, the difference was not significant. This result shows a higher reproducibility for our proposed method. We further repeated this experiment with only six measurements in each subset to simulate clinical scans. In this experiment, on all 41 tracts $\Delta \text{FA}_{\text{tract\_}i}$ was smaller for our method than for TBSS; 22 of these differences were statistically significant at $p=0.01$. We also plotted the profile of FA on the centerline of various tracts. The profiles computed with our method showed higher consistency and reproducibility.

\textbf{Robustness to data perturbation.} For each of the 35 test subjects from the HCP dataset, we added random Rician noise with a signal-to-noise-ratio of 20dB to the dMRI measurements. We applied our method and TBSS on these noisy scans as well as on the original scans and computed the difference in mean FA for tract $i$ and subject $j$ as $\Delta \text{FA}_{\text{tract\_}i,j}= \big| \text{FA}_{\text{tract\_}i,j}^{\text{noisy}} - \text{FA}_{\text{tract\_}i,j}^{\text{original}} \big|$. We denote the average of $\Delta \text{FA}_{\text{tract\_}i,j}$ across all subjects with $\Delta \text{FA}_{\text{tract\_}i}$. On all 41 tracts, $\Delta \text{FA}_{\text{tract\_}i}$ was smaller for our method than for TBSS. Paired t-tests showed that 28 of these differences were statistically significant at $p=0.01$. This result shows that our method is more robust to data perturbations.

\textbf{Analysis of the impact of gender and age with an independent dataset.} We performed additional experiments on the Calgary Preschool dataset. With the help of a semi-manual tract segmentation technique \cite{reynolds2019white}, this dataset had been previously used to assess the impact of age and gender on the micro-structural changes on 10 major white matter tracts \cite{reynolds2019global}. We applied our proposed method and TBSS on this dataset to determine the impact of gender and age on FA and Mean Diffusivity (MD), and compared the results with \cite{reynolds2019global}. This experiment showed that, compared with TBSS, the results of our method were consistently closer to the results of \cite{reynolds2019global}. For example, our method showed a significantly ($p<0.01$) higher FA on pyramidal tracts and higher MD on the superior longitudinal fasciculus for boys compared with girls, which was consistent with the results of \cite{reynolds2019global}; TBSS did not reveal these differences. We also performed a reproducibility experiment with this dataset. This dataset included 75 subjects (30 female) aged 3 years and 73 subjects (34 female) aged 6 years. We divided these into \{Subset 1: 37 (15 female) aged 3 \& 37 (17 female) aged 6 \} and \{Subset 2: 38 (15 female) aged 3 \& 36 (17 female) aged 6\}. We computed the change in FA and MD for each tract using the two subsets. Ideally, the change computed with the two subsets should be equal. In terms of FA for 35 of the tracts and in terms of MD for all 41 tracts the difference between the changes computed using the two subsets was smaller with our method than with TBSS. This experiment shows higher reproducibility for the proposed method.

\section{Conclusions}

\emph{The potential impact of the proposed method cannot be overstated.} Tract-specific studies are increasingly used to answer critical research questions in medicine and neuroscience. However, the shortcomings of existing computational methods have seriously limited the accuracy and reproducibility of these studies. Our experiments showed that the method proposed in this paper offers superior reproducibility and robustness to data perturbations. Experiments with the Calgary Preschool dataset further showed that our method has high generalizability. Therefore, our proposed method can have a significant impact by enhancing the accuracy, reliability, and reproducibility of the results of tract-specific studies.

\section*{Acknowledgment}

This research was supported in part by the National Institute of Biomedical Imaging and Bioengineering, the National Institute of Neurological Disorders and Stroke, and Eunice Kennedy Shriver National Institute of Child Health and Human Development of the National Institutes of Health (NIH) under award numbers R01HD110772, R01NS128281, R01NS106030, R01EB018988, R01EB031849, R01EB032366, and R01HD109395. This research was also partly supported by NVIDIA Corporation and utilized NVIDIA RTX A6000 and RTX A5000 GPUs. The content of this publication is solely the responsibility of the authors and does not necessarily represent the official views of the NIH or NVIDIA. 
This work was also supported by the Swiss National Science Foundation (project 205321-182602). We acknowledge access to the facilities and expertise of the CIBM Center for Biomedical Imaging, a Swiss research center of excellence founded and supported by Lausanne University Hospital (CHUV), University of Lausanne (UNIL), Ecole polytechnique fédérale de Lausanne (EPFL), University of Geneva (UNIGE) and Geneva University Hospitals (HUG).

\clearpage

\bibliographystyle{splncs04}
\bibliography{davoodreferences}

\end{document}